FRIEDRICH-ALEXANDER
UNIVERSITÄT
ERLANGEN-NÜRNBERG
FACHBEREICH WIRTSCHAFTS-
WISSENSCHAFTEN

# Retail Central Bank Digital Currencies, Disintermediation and Financial Privacy: The Case of the Bahamian Sand Dollar

Manuscript

Author: Kilian Wenker

# Abstract

The fast-growing, market-driven demand for cryptocurrencies worries central banks, as their monetary policy could be completely undermined. Central bank digital currencies (CBDCs) could offer a solution, yet our understanding of their design and consequences is in its infancy. This non-technical paper examines how The Bahamas has designed the Sand Dollar, the first real-world instance of a retail CBDC. It contrasts the Sand Dollar with definition-based specifications. I then develop a scenario analysis to illustrate commercial bank risks. In this process, the central bank becomes a deposit monopolist, leading to high funding risks, disintermediation risks, and solvency risks for the commercial banking sector. I argue that restrictions and caps will be the new specifications of a regulatory framework for CBDCs if disintermediation in the banking sector is to be prevented. I identify the anonymity of CBDCs as a comparative disadvantage that will affect their adoption. These findings provide insight into governance problems facing central banks, and coherently lead to the design of the Sand Dollar. I conclude by suggesting that combating cryptocurrencies is a task that cannot be solved by a CBDC.

# Index

# List of illustrations

# List of tables

# List of abbreviations

| | |
|---|---|
| AFI | authorized financial institution |
| AML/CFT | anti-money laundering and countering the financing of terrorism |
| App | azpplication program |
| ATA | anti-tax avoidance |
| Bafin | Bundesanstalt für Finanzdienstleistungsaufsicht |
| CB | central bank |
| CBDC | central bank digital currency |
| CBOB | Central Bank of The Bahamas |
| DLT | distributed ledger technology |
| E-money | electronic money |
| FATF | Financial Action Task Force (on Money Laundering) |
| IOU | abbreviated from the phrase “I owe you”, a document acknowledging debt |
| KYC | know your customer |
| PSP | payment service provider |
| rCBDC | retail central bank digital currency |

# 1. Introduction

Central Bank Digital Currencies (CBDCs) refer to legal tender in digital form. Their introduction could radically change the banking sector, and it is already on its way. This introduction will present the topic and its relevance, position my approach and objectives, and give an overview of the paper's structure.

Monetary stability is a major concern of central banks (CBs). Because of the long-term relationship between monetary growth and inflation, a CB tracks the growth of monetary aggregates. This is where the growth of private cryptocurrencies becomes an issue. How can a CB track and control the growth of the money supply when monetary functions are taken over by cryptocurrencies that are intentionally obfuscated and largely thrive outside of the national legal framework? Using a retail CBDC (rCBDC) as legal tender might offer a solution.

However, a rCBDC could compete with payment accounts at commercial banks, especially if it bears interest. The core business of commercial banks might break: The latter provide savings accounts, facilitate payments and provide lending — but without funding, there is no lending. If a commercial bank loses most or all of its deposits, how can it keep up its balance-sheet to sustain lending to businesses?

Moreover, banks interface the state with the economy, providing a certain degree of anonymity. Any CBDC would provide governments with a technical framework enabling complete control. How can one balance privacy and the tracing of illicit transactions in a CBDC setting?

This paper aims to provide some clarity on two challenges of CBDCs. First, what impact might a CBDC have on commercial bank funding risks and banking stability? Could a CBDC cause financial disintermediation? Second, a CBDC may open up new policy options, such as truly full government control on payments. CBDCs could become "panopticons for the state to control citizens: think of instant e-fines for bad behavior" [1]. This paper focuses on the opposite question: could a CDBC, intended to serve as the digital equivalent of cash, achieve the anonymity of cash or cryptocurrencies?

A case-study approach was taken to gain an understanding of a CDBC in a real-world setting. The Bahamas provides a very insightful account of a CBDC. First, The Bahamas is applying practical solutions to address the risks and theoretical difficulties noted above. Second, there is no significant political burden that would affect the design of its rCBDC (unlike, for example, the digital ruble or the eYuan).

The query TITLE-ABS-KEY ( central AND bank AND digital AND currency ) yielded scarce results from the Scopus database, although the number of publications on this topic has been steadily increasing since 2018 (see Figure 1). To expand this thin base of scholarly articles, this paper draws on a variety of studies outside of the Scopus database.

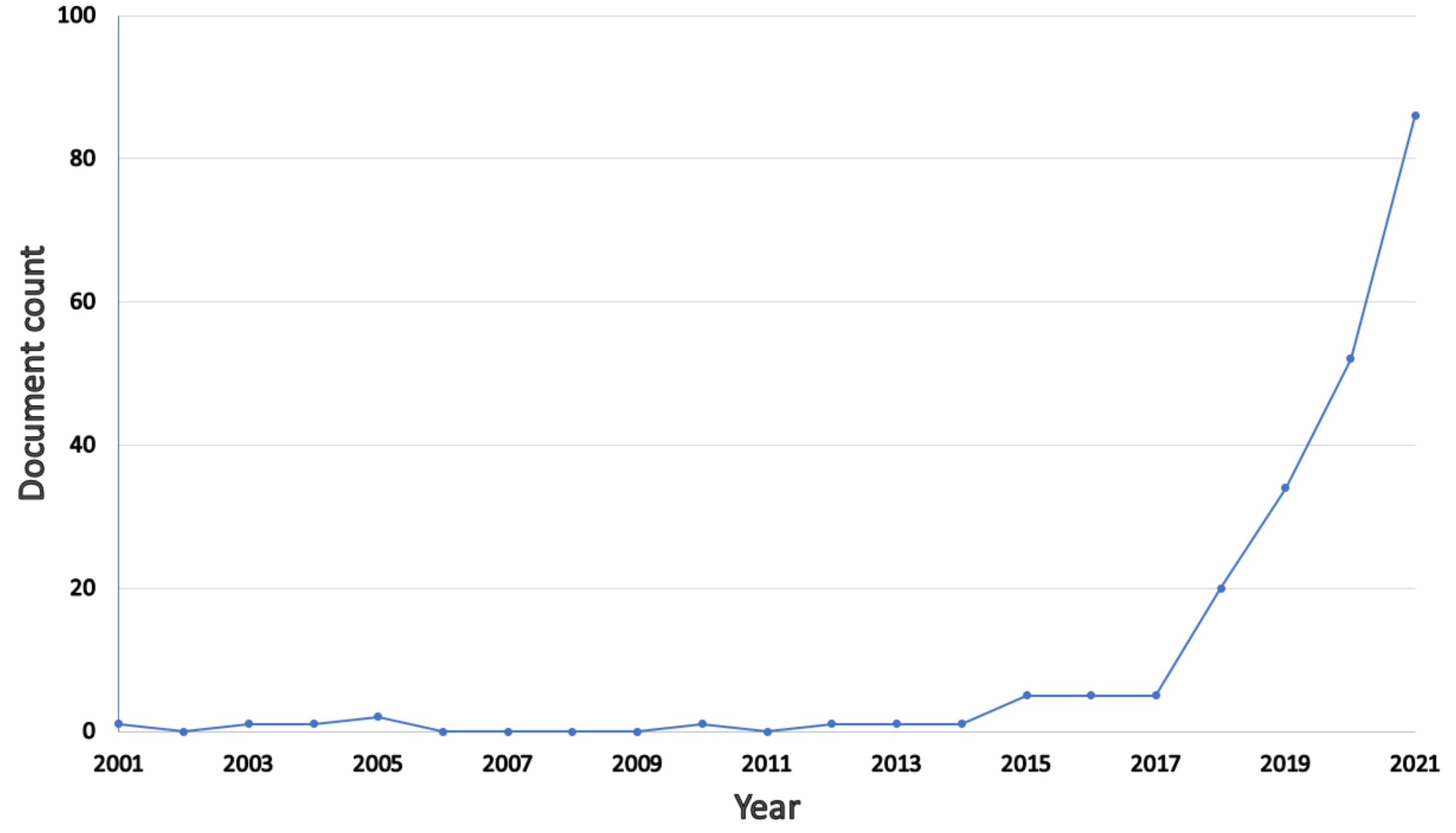

**Figure 1: Number of academic documents per year**

On a side note, some questions cannot be resolved without explaining the functional and technical design of a CBDC.

This paper is organized as follows. Section 2 presents the theoretical background. Section 3 deals with the CBDC of The Bahamas to provide a real-world instance and provides an overview of the design features of CBDCs. Many technical details that are interesting in their own right are beyond the scope of this paper. Sections 4 and 5 explore disintermediation and privacy. Finally, the paper is concluded in Section 6.

# 2. Theoretical background

This section addresses the concepts of currency, money, and cryptocurrency, and outlines the theoretical framework of competitive money supply.

## 2.1 Currencies and monetary systems

Despite its common usage, the term currency is rarely defined. In general, the state is the monopolistic supplier of the currency as the official means of payment [2-4]. Currency is therefore a "creature of the state" and a simple public monopoly [5]. A functional currency is legal tender, in circulation, and customarily used as a medium of exchange in the country of issuance [6]. The term currency is used in this paper to refer to the official means of payment of a state/currency union, issued by sovereign entities, either in physical form and designated as legal tender, or in electronic form and legally recognized. This definition is loosely based on [4] but softens the status of legal tender for electronic variants.

A CB often has the derivative power to issue currency and uses monetary policy to stabilize either economic growth or inflation or both. It serves as the bank of the commercial banks in its country and runs the monetary policy which in turn affects the stability of the currency [7].

Money is a broader concept than currency. In this paper, money is referred to as anything that performs the following three functions: Store of value, unit of account, and medium of exchange [8–11]. Even the simple "Monopoly money," an in-game currency used in the popular board game, meets this definition within the game.

E-money (or digital money) refers to money in purely digital form, for example, money held by individuals in bank accounts or by commercial banks in deposits at the CB [11]. LHV Pank, an Estonian commercial bank, was the first bank in the world to experiment with programmable money when it issued €100,000 worth of cryptographically protected certificates of deposit, denominated in euros [12]. LHV's Cuber Wallet app enables users to send and receive euros instantly, using a distributed ledger technology (DLT). That technology will be explained in the next subsection.

## 2.2 Cryptocurrencies

A cryptocurrency is a permanent, digital database designed to work as a medium of exchange. As stated in [9], that database records peer-to-peer transactions one after the other in a continuous ledger, permanently, so that latter can only be accessed and updated. This ledger is spread across multiple websites, countries, institutions, and users, hence the name DLT. The security and accuracy of the assets stored in the ledger is upheld cryptographically strong through the use of "keys" and signatures to control who can do what within the shared ledger.

Fraudulent transactions in the form of double-spending attacks — where users spend the same money at least twice — can be a problem in decentralized systems. Sophisticated or larger networks prevent double spending by implementing a confirmation mechanism and maintaining a common, universal ledger system, and by setting high hash rates [12, 13].

Despite their name, most cryptocurrencies do not meet the criteria of a currency because they are not legal tender (see Section 3.1 for a counterexample). They also do not function as money because they do not fulfill all three functions of money. For example, a medium of exchange requires general acceptance. Some authors [9–11, 14] argue that store of value requires less price volatility. On the other hand, if one ac-cepts that Monopoly money serves as money for the restricted group of Monopoly players, then one must also accept that a cryptocurrency like Bitcoin, the most widely used at this moment, acts as money for the community of Bitcoin users. To date, crypto-assets might be a more appropriate term [14–16], especially since many users hold Bitcoins as an investment rather than use it to complete transactions [8, 9].

Stablecoins, a special category of e-money, solve the problem of unstable purchasing power caused by high exchange rate volatility by tethering their value to a currency, commodity, or basket of assets [8, 11, 17]; see also LHV's Cuber in subsection 2.1 for an example.

### 2.3 Monetary competition

Demand for cash is decreasing while the use of e-monies is increasing [10, 11, 17–20]. From a systems design perspective, the growth and fall of payment methods will lead to situations in which outcomes become a part of competitive thinking. The same is true for currencies and monies, where a key paradigm is the assumption that "good money", especially money enjoying consumer trust, will squeeze out the weaker monies (see [2, 21] for a theoretical and historical perspective). However, competition between currencies is not the only possible approach in an environment of currency plurality; complementarity must also be considered [22].

A major concern of competitive money supply is financial stability and the loss of consumer confidence. In uncertain times, there could be a shift from bank deposits to cash, i.e., from digital money to physical money. According to [11], this was the case in 2008, when the flight from bank deposits peaked after the collapse of Lehman Brothers, and continued during the 2010-13 sovereign debt crisis. If there had been a risk-free, digital version of household cash back then, bank customers would not have to form long lines to withdraw cash at bank branches or ATMs in this situation — they could do so conveniently on their cell phones, from digital money to digital money.

Another venue for the flight to safety could be another currency. Currency substitution (dollarization, euroization) is common in countries where confidence in the domestic currency is waning. Practitioners contemplate "a currency crisis in an emerging country a decade from now, when people and businesses can choose to make and collect payments in yuan instead of in their local currency, in the time that it takes to generate a QR code on a phone" [23]. The last part is the import aspect. E-monies increase the speed of exchange and ease of use. Could cryptocurrencies actually accomplish such a task? Amount and confidence will be crucial. Nearly 17,000 crypto-assets have a market capitalization of two trillion U.S. dollars now, with Bitcoin dominating the market at about 40 percent [24]. This scale implies confidence and has raised concerns that cryptocurrencies could influence national monetary policy [7, 10, 12, 25].

A simplified illustration of typical forms of competing money can be found in Table 1.

| | Physical money | Digital money |
|---|---|---|
| **Legal tender (or the nearest equivalent)** | Banknotes and coins | **CBDC**s |
| **Regulated to a lower standard, not legal tender** | nil (some IOUs such as German "Notgeld" come close) | Accounts at AFIs, a Visa-card, PayPal, Alipay, LHV's Cuber etc. |
| **Unregulated** | Commodity money such as gold coins; local coupons; etc. | Cryptocurrencies like Bitcoin, Ethereum, etc. |

**Table 1: The different types of competing money**

This table does not include physical and electronic play money, as it cannot seriously be said to compete with an official currency or real-world money. Accounts at AFIs is a very general formulation that can include, for example, time deposits at a commercial bank, reserves at a CB or balances at a payment service provider (PSP).

# 3. CBDCs in general and the case of The Bahamas

This section will present the CBDC currently deployed in The Bahamas and then explore the CBDC concept in general.

## 3.1 The Bahamian Sand Dollar

On 20 October 2020, The Commonwealth of The Bahamas (The Bahamas) became the first country to deploy a nationwide CBDC by introducing the Sand Dollar. The Central Bank of The Bahamas (CBOB) had first piloted a digital version of the Bahamian dollar in the Exuma district starting 27 December 2019, and had it extended to the Abacos a few months later [26, 27].

The Sand Dollar is pegged 1-for-1 to the Bahamian dollar, the currency of The Bahamas, which is in turn pegged 1-for-1 to the U.S. dollar. Two-thirds of all jobs in The Bahamas are attributable to tourism, and since about 80% of tourists come from North America, the easy conversion rate makes many merchants accept U.S. dollar bills [19, 28]. The Sand Dollar is a direct liability of the CBOB, backed by the foreign reserves [29].

In the first few weeks through the end of 2020, the CBOB issued limited amounts of Sand Dollars to Authorized Financial Institutions (AFIs) and had a total worth of 130,000 Sand Dollars in circulation at year-end [30]. Since then, the total worth of Sand Dollars in circulation has increased to 302,785.04 [31]. About 28,000 eWallets use the Sand Dollar [31], this amounts to roughly 7% of the country's population.

**The Sand Dollar architecture.** The Sand Dollar requires a technical platform to process payment transactions. NZIA Ltd. is the technical services provider to which CBOB outsources most of these technical services. Since the prepaid cards or digital wallets contain CB money and are based on DLT, the transactions can be processed directly between the eWallets of the payer and the payee. Figure 2 illustrates the basic operation.

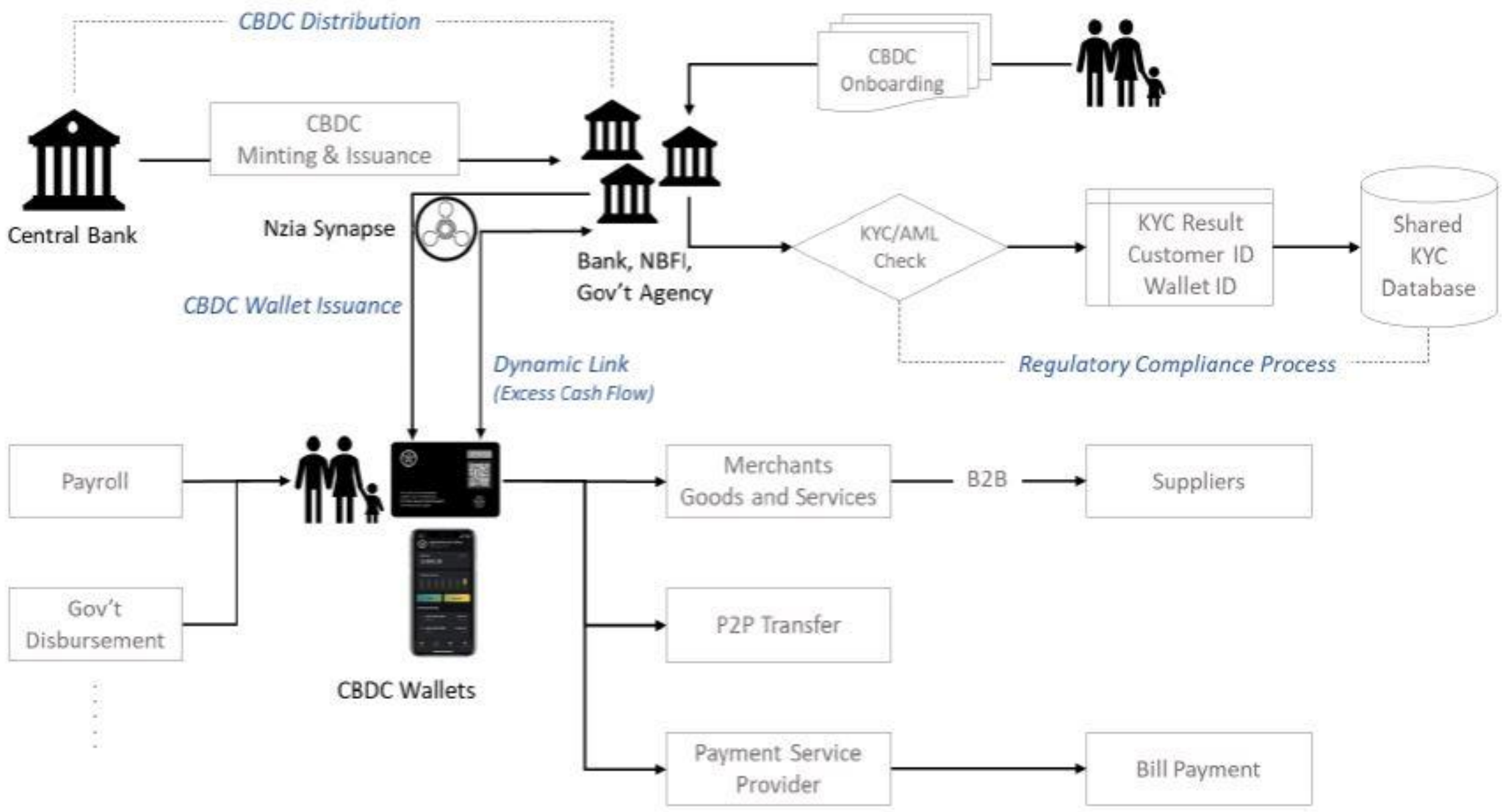

**Figure 2: The Sand Dollar architecture**

**(source: [27], p. 21)**

**Accessibility.** The Sand Dollar is a wholesale CBDC for "settlements at the inter-bank level, akin to clearing house transactions" [29] as well as a rCBDC [27]. In its original form, consumers could pay with Sand Dollars only through an app at specific merchants. With a prepaid card recently introduced by Mastercard Inc., Bahamian consumers can pay with the Sand Dollar anywhere "Mastercard" is accepted around the world [32]. Consumers can choose between a Tier I eWallet with $500 holding limit, with a $1,500 monthly transaction limit, and a Tier II eWallet with a $8,000 holding limit, with a $10,000 monthly transaction limit [27, 33]. For residents, the intended outcome of Project Sand Dollar is that they can all use a CBDC with an experience and convenience — legally and otherwise — that resembles cash [27]. This includes offline functionality that is not yet fully developed, e.g., if inter-island communications are interrupted, built-in safeguards should allow users to pay a pre-determined dollar amount, and eWallets should be updated once communications with the network are restored [27].  As the governor of the Bahamian CB explains, in some cases people literally text money from one person to another [34].

How does a CBDC differentiate itself from private PSPs like Apple Pay or PayPal? The Sand Dollar does not compete with them, but provides the foundation that these payment services can use as an interface, and, in fact, more spending occurs across platforms than at retail [35]. Two people are no longer hindered by using different PSPs; they can always use Sand Dollars Moreover, a private PSP could restrict or block an account comparatively easily, while the standard of security for the Sand Dollar user must be much higher (keyword here: legal tender).

**Objectives.** The Bahamas is a financial center, but its natural environment makes it difficult for many residents to access financial services. Geographically, The Bahamas is an archipelago consisting of 700 islands scattered across a vast expanse of ocean, with a 93% penetration for mobile devices; about 96% of surveyed Exumians own mobile devices [19, 26, 27]. As the governor of the CBOB points out: “One of the limitations of being an island archipelago is that even if you have a bank, depending on where you live, you probably have to take a trip to Nassau and go to the bank. Some people do that, as ridiculous as it sounds. If you live in some of the remote communities, it’s a half-day or a full-day event to get to the bank” [36]. Thus, the fundamental advantage of the Sand Dollar might be that it is easier to distribute than cash, especially among the underbanked and unbanked. The CBOB stresses the following objectives: To provide comprehensive, non-discriminatory access to payment systems; to increase the efficiency of Bahamian payment systems; and to strengthen efforts against money laundering, counterfeiting, and other illicit purposes by reducing the negative impact of cash use [37].

## 3.2 CBDCs in general

The real-world example of The Bahamas reveals that CBDCs are digital fiat — the Sand Dollar is an extension of the Bahamian dollar. The 1-for-1 convertibility is necessary to maintain the function of unity of account of the currency. Anything else would lead to an exchange rate between different types of CB money and break the unity of the currency [14].

There is no clear single definition of CBDCs [4, 14, 18]. Throughout this paper, the term CBDC will refer to a CB liability in digital form, denominated in the official national currency (like the Bahamian dollar) or an equivalent at a fixed conversion rate (like the Sand Dollar), issued and regulated by a sovereign entity, and intended as an electronic substitute for cash in daily transactions.

This definition avoids the term legal tender that could be misunderstood to force vendors and other creditors to invest in potentially expensive equipment because they may be compelled to accept a rCBDC due to its legal tender status [4, 25]. This line of reasoning is comprehensible, but I do not endorse it as limits and caps are not uncommon in legal tender status (for example, banknotes are legal tender in Canada but there is no legal duty for vendors to accept them [18]; EU law states in Council Regulation (EC) No 974/98 of 3 May 1998 that no party shall be obliged to accept more than 50 coins in any single payment; the Bahamian Sand Dollar is legal tender with all of its constraints and ceilings mentioned in this paper; see [25] for several divergent accounts of legal tender).

It is important to note that this is a CB liability, not a private company liability. If private companies issue a similar liability, then it is not a CBDC, but a stablecoin such as LHV's Cuber.

CB liability is not only a legal construct, but also has a financial aspect. The latter can mean that there are accounts directly at the CB for everyone. But it can also mean, for example, that there are segregated reserves of financial institutions, such as commercial banks and PSPs, with the CB while individuals have the legal equivalent of an account at the CB. A variant of the first option was chosen for the Sand Dollar; "Sand Dollar in circulation" is now an official line item on the CBOB's balance sheet.

The intention to replace cash is outlined, as the definition could otherwise refer exclusively to the demand deposits (also known as reserves or settlement balances) that already exist in real terms at CBs (see [18] for a different view). The volume and trend of cash use in a given country will ultimately determine demand for CBDCs [25]. The intention to replace cash leverages very low transaction fees, ideally zero, and very small requirements for technological investments (like an app on a smartphone). Cash-like features could be interpreted as strong user privacy protections for low-value transactions (see Section 5) and interest-free deposits. In the case of the Sand Dollar, this feature leads indeed to low KYC requirements; no official ID is required for the Tier 1 eWallet. Finally, the intention to replace cash was a reason for rejecting interest on the Sand Dollar [27].

Although the intended use is part of the definition, the user group is not addressed: The restriction to consumers has the potential to exclude wholesale CBDCs.

I will briefly discuss two design choices, the underlying technology and the overall accessibility, as these will play a role in the following sections.

**Underlying technology.** Token-based e-monies share outward similarities in their technology, such as the use of DLT, while account-based e-monies require an intermediary, usually a bank, that accepts deposits and keeps records in a ledger [21]. A CBDC could use both technologies, as shown in Figure 3. The centralized CBDC account option with a central validator in Figure 3 potentially precludes peer-to-peer transfers (which is why I did not include the peer-to-peer feature in the definition of a CBDC). And yet, even a DLT-based token CBDC, the last option in Figure 3, also offers an account, but the account is managed on a decentralized basis. Some authors think that a CB will rather abstain from using DLT, e.g. because of the finality of payments in DLT [21]. Others advocate the opposite [38]. It is obvious that technology does not define CBDC. But the chosen technology has implications for privacy and tracking.

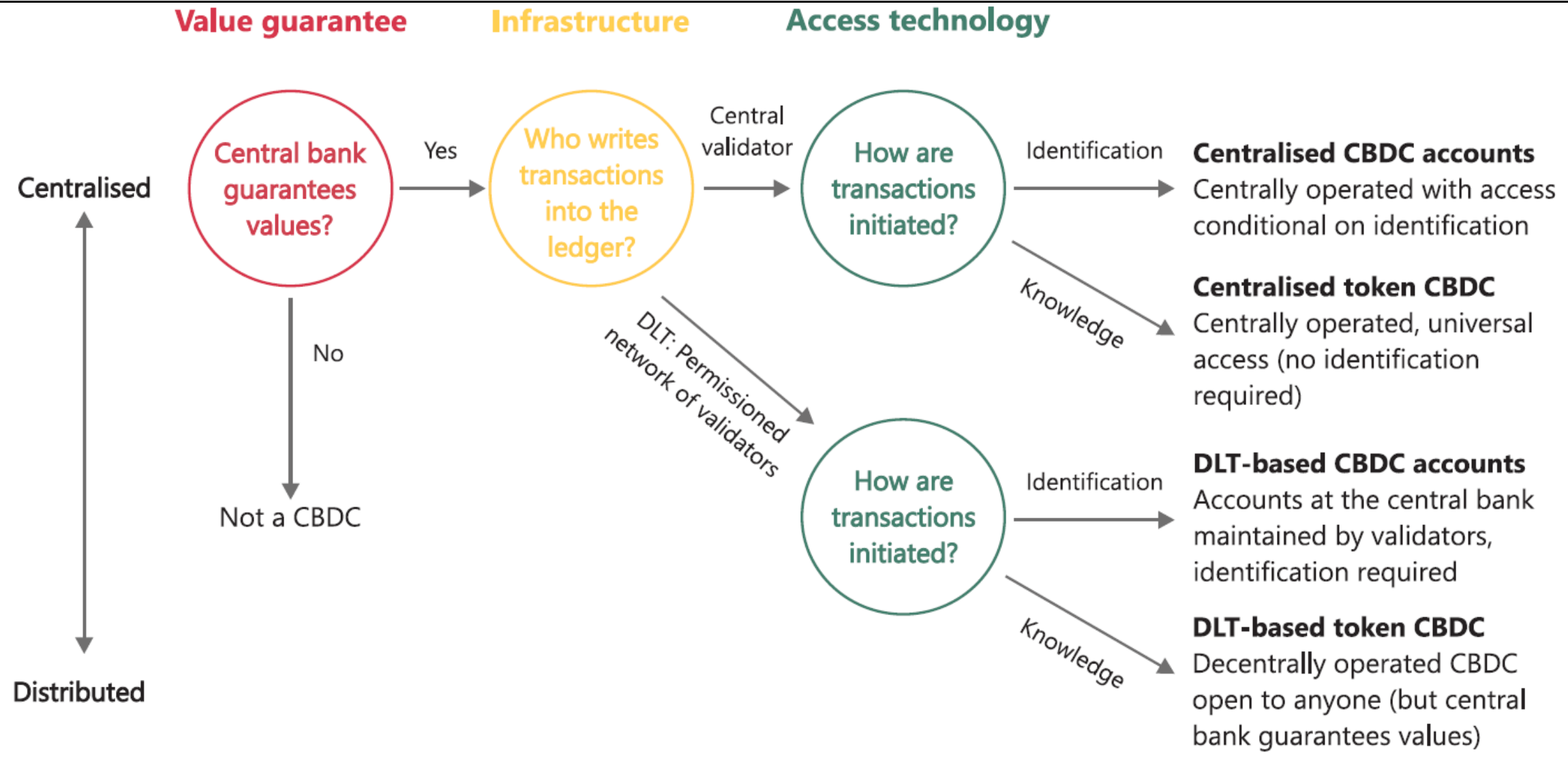

**Figure 3: Four possible combinations of a CBDC infrastructure (source: [39], p. 92)**

**Accessibility.** CBDCs can be divided by accessibility into wholesale CBDCs, where the network participants are financial institutions that already have access to the CB's balance sheet, and rCBDCs, which are available to general users such as businesses and households [21]. The latter are also referred to as direct CBDCs.

The main rationale behind wholesale-only CBDCs are either a better domestic wholesale, real-time gross settlement system in emerging market economies or increased efficiency for cross-border payments in advanced economies [19]. These pure wholesale CBDCs are not included in the definition of CBDCs in this paper, as those purposes would substantially lessen CBDCs to the introduction of a more efficient technology (increasing transaction speed and decreasing transaction costs) for existing CB settlements, maybe making them legal tender (they are, of course, already a CB liability). But there would be no access to the CB for individuals and businesses in a purely wholesale CBDC.

The Sand Dollar disrupts that architecture by creating a two-tier system with AFIs (see previous subsection) handling retail payments while each holder of the Sand Dollar maintains direct claims on the CB and legally has the equivalent of accounts with the CB [29]. This results in the CB holding the retail balances and is referred to as a hybrid

CBDC [39]. Banks and PSPs act as agents of the CB, which means that retail customers' balances with the CB are not shown on the AFIs' balance sheets.

Another notable mixed form is the "synthetic" or "indirect" CBDC. It is the equivalent of narrow-bank money [4, 39]. In this case, commercial banks issue e-money comparable to a stablecoin. Wholesale reserves at the CB act as full reserve for the commercial banks' e-money, but the retail customer has no direct claim on the CB. The major CBs do not consider such a design as CBDC [40], and my definition excludes it as well, since it requires CB liability, CB issuance, and legal tender (or equivalence).

One more way to restrict access is to impose limits or caps. The CBOB has introduced a rCBDC that is limited to Bahamian residents and sets a cap on transfers and account balances per holder — it is meant to replace cash, not bank accounts. The emphasis on domestic use will expand as tourists gain access to the Sand Dollar.

# 4. Commercial bank funding risks in a cashless economy

This section discusses risks to commercial banks that arise when a rCBDC increasingly wins monetary competition on deposits at commercial banks. To this end, I will provide a scenario analysis that does not aim to make predictions, but to provide alternative pictures of the future evolution of the CBDC environment. The three scenarios are illustrative in nature to provoke thinking. They are not detailed blueprints.

The basic assumption of the scenario analysis is a rCBDC that offers a free, low-risk, interest-bearing account at the CB, offering fast payments without limits, in a cashless economy. Ceteris paribus conditions include fractional reserve banking and CB funding of commercial banks (see [20] for a discussion of the impact of CBDCs on this ceteris paribus clause). These ceteris paribus conditions rule out a greater role for CBs in financial intermediation from the outset. It is tantamount to CBs watching and not funding commercial banks when, for example, commercial banks' customer deposits melt down. In the scenario analysis, it is assumed that the substitution between bank deposits and rCBDCs is completely unrestricted. The analysis proceeds in three steps.

## 4.1 High-cost funding risk: less profitability for commercial banks

The basic assumption of the first scenario is that commercial banks lose their demand deposits entirely to the risk-free deposits with the CB. For the commercial banks, less demand deposits automatically mean less funding, e.g., to finance loans with short-term liabilities. Various reactions by the commercial banks are conceivable.

The first conceivable option is simply to do nothing, for whatever reason. This would lead to a contraction of the commercial banks' balance sheets and to less profitability.

Active countermeasures, however, could be taken via the liabilities side of the balance sheet. If demand deposits disappear, the commercial bank can replenish the asset side either by refinancing itself via the wholesale funding markets or by increasing longer-term deposits in the retail sector. However, both options would be more costly. Longer-term deposits would have to offer higher interest rates in order to attract more deposits and would thus be more costly. Increased reliance on market funding makes banks more vulnerable to unexpected changes in market conditions [20] and should be expected to be more costly than accounts at commercial banks.

On the assets side, there are few opportunities to actively counteract this: More investments on the asset side are only feasible if more funds are available on the liability side. Nevertheless, commercial banks could try out four strategies. First, they could invest in riskier assets with higher yields. But this would not increase funding, would leave banks less stable and, in this respect, does not seem very likely — if it were that easy, commercial banks would have been pursuing higher yields at the same level of risk long ago. Second, they could try to charge higher interest rates on loans. In theory, this might improve profitability; in practice, they would lose market share, and again, it would not generate higher funds for them. A third strategy would be to divest themselves of certain assets and instead put more funds into loans to households and businesses. Putting less money into debt securities would be a simple example, but it would create several problems, such as less financial robustness (investment in debt securities aims to manage interest rate and liquidity risk) and usually nowhere near enough volume to offset the loss of deposits. A fourth and final option might be to link lending to deposit and payments business, thereby making deposits mandatory. But no business or household would want to be forced to do all its business at a single bank. I don't think this strategy is easy to implement, and it would have to be accompanied by very favorable conditions for payment accounts and would reduce profitability.

Roughly summarized, the restructuring of liabilities could theoretically absorb the loss of demand deposits, but at an increasing cost, and in turn the supply of credit would decline due to the pass-through of costs to the credit market (I assume that interest rates are exogenous). Assets, on the other hand, cannot compensate for this loss, and measures on the asset side would tend to worsen the stability and liquidity of commercial banks.

An exception is conceivable with regard to banks' funding costs for maturity transformation. If, for example, the decline in the operating costs of payment accounts and the rise in the interest rate on, say, term deposits cancel each other out, there is no significant impact on the supply of loans and bank profitability. However, this would leave open the question of what customers now use for their daily payments when they transfer their money from demand deposits to term deposits instead of CBDC eWallets. In other words, could rCBDCs work at all if no one uses them (but the basic assumption of the scenario analysis is that rCBDCs are a very effective substitute for traditional demand deposits).

The previous paragraphs have stressed the impact of an rCBDC from the perspective of a commercial bank's balance sheet, yet one off-balance sheet issue should be highlighted. A rCBDC could severely restrict supply in the interbank lending market if bank deposits are shifted to the CBDC. This would amplify the impact of higher wholesale funding costs.

**Why should this happen?** Briefly take the point of view of an individual customer: The appeal of a cost-free, risk-free, and instant payment account is conspicuous. Combine this attractiveness with the inertia of some banks. Bahamian banks are cautious about the Sand Dollar. The six AFIs initially approved after satisfactory security clearance were all PSPs [30]. By July 2021, nine PSPs and finally two banks had been approved [41]. Mastercard Inc. is an American multinational financial services company, focused on electronic payments, rather than a traditional bank. But it was the first multinational to add the Sand Dollar to its product portfolio, well ahead of commercial banks and in collaboration with Island Pay, a local PSP [32]. Next step: The CBOB plans to eliminate all use of domestic cheques by the end of 2024 [42], another bank-related means of payment.

The trigger for this scenario does not have to come from weighing economic benefits; it may come from the political environment. In 2018, the full money initiative ("*Vollgeld-Initiative*") forced a referendum that would have given the Swiss CB a monopoly on issuing demand deposits in Switzerland [43]. There are similar initiatives in other countries.

### 4.2 Disintermediation risk: new business models become inevitable

In addition to the first scenario, the second scenario assumes a crowding out of medium- to long-term debt instruments of commercial banks because, for example, individuals or businesses prefer to invest in crypto assets rather than in term deposits, bonds or other longer-term debt instruments of commercial banks. Or a new generation of interest-bearing cryptocurrencies makes debt securities become unappealing assets. Or an inverted yield curve grants higher yields for CBDCs than for long-term commercial bank instruments (assuming usually the same credit risk profile).

At the end of the process, only the equity of commercial banks remains to funnel loans. Turned positively, one could therefore say that if the CB monopolistically takes over all deposits, then bank runs are technically no longer possible. New business models would emerge, such as banks servicing only the asset side of their balance sheet because they lack retail and wholesale funding entirely, and PSPs and CBDCs would take over payment services entirely. Any residual deposits with commercial banks could at best be used to fund banking operations, not lending. Complete disintermediation of banks has been achieved. Investment banks could flourish in this environment.

By analogy with the reasoning in Section 4.1, the supply of loans to the real economy would either tighten sharply or lending conditions would deteriorate drastically as commercial banks would have to use more expensive funds. This, in turn, would break up the loan market and lead directly to the next scenario in Section 4.3.

**Why should this happen?** Because technology has disrupted so many industries, its impact on banking may seem like another example of a cumbersome, uncompetitive business made obsolete by savvy technology companies [23]. Investors have already invested two trillion U.S. dollars in crypto-assets [24]. There are, moreover, historical examples of how CBs strongly dominate the market for deposits (see [44] for the Bank of Spain in 1874-1913).

### 4.3 Solvency risk: bank failure and bank run

The ultimate risk for commercial banks is, of course, that their very existence is threatened. Suppose an individual wants to buy a new car and the car manufacturer offers financing with a stablecoin, which in turn is linked to the car via DLT. The principle is simple: if the customer does not pay his monthly installments, he cannot unlock his car, its doors remain locked. As the core of commercial banks' traditional business model — taking short term deposits and funding longer term loans — fails, the result in terms of market structure will be that one commercial bank after another will have to be resolved if the commercial banks as a whole do not succeed in reinventing their business model. CBDCs and DLT would have been only the forerunners of this development.

Solvency risk may result from the fading business model, but it could also be rooted in consumer confidence. A bank run would hardly be possible in the last scenario, as the liability side of the balance sheet represents 100% equity at the end of the disintermediation risk scenario. Nevertheless, I assign bank runs to the third scenario, since they are part of solvency risk.

A bank run would occur much more quickly in a digital world without restrictions; a single wire transfer would be enough to turn the deposit into a risk-free rCBDC. The CBDC would be a flight-to-safety instrument whose very existence could be destabilizing. A mixed CBDC variant, e.g., a wholesale variant with unrestricted retail accounts at commercial banks would not be able to curtail this "instrument". Moreover, a deposit guarantee scheme cannot be considered a stabilizing factor in this scenario, as recent history has shown that a deposit guarantee scheme can be quickly adjusted in a financial crisis [20]. This amounts to saying that a rCBDC could abolish implicit and explicit guarantees on commercial bank money [20].

**Why should this happen?** This seems rather unlikely at the moment. In the second scenario at the latest, the large commercial banks would presumably buy up PSPs, and replenish their own liability side with the PSPs' deposits. A commercial bank's expertise and experience in credit assessment could be difficult to copy by technology. Yet, there are already small-scale examples of tokenization of SME loans used to trade loans for small businesses, approved by the Bafin, the regulator for national financial markets in Germany [45].

The CBOB has created various restrictions to prevent these scenarios from coming to fruition: Restrictions on users, restrictions on amounts, approval requirements for AFIs, no interest on Sand Dollars, etc. And that extends to the risk of bank runs: The CBOB "will deploy circuit breakers, if necessary, to prevent systemic instances of failures or runs on bank liquidity" [27]. This leads to another corollary: Financial stability analysis often focuses on issuers, be they commercial banks or PSPs, in particular on their capital adequacy, stress testing and market liquidity risk (think Basel III), but constraints and caps will complement financial regulation in the future.

### 4.4 Results

This scenario analysis has its limitations. It hides the impact on CBs from the outset (potentially larger CB footprint in the financial system, higher exposure to credit risks, more power to the CB, etc.), but clearly shows the risks to commercial banks.

**Interest payments.** The scenarios would work in much the same way if there were no interest payments on CBDCs, although the substitution would be less aggressive and the change more lenient. For example, an interest-free CDBC could be more attractive than interest-bearing commercial bank deposits if the risk assessment is markedly different. Moreover, an interest-bearing CBDC cannot be ruled out for two reasons. First, the CB needs to provide an additional incentive for the use of its CBDC, otherwise it will not be more attractive than private solutions such as Alipay, Bitcoin or credit cards, etc. Indeed, the CBOB must make efforts to convince citizens and AFIs to use the Sand Dollar. Second, a CBDC interest rate could serve as the main tool for controlling monetary policy.

**Cash.** Users trade one characteristic for another when deciding which types of money to hold in their portfolio. The existence of three regulated types of money (see Table 1) theoretically means that none of them dominates in all features (such as interest rate, issuer, risk, insurance of payments, ease of use, etc.). The basic assumption that cash no longer exists reduces the portfolio choices of households and non-financial businesses to commercial bank money and CBDC. This is consistent with CBDC's purpose of replacing cash. However, the reality is much more heterogeneous and there is no uniformity of money or currency. A deposit at a vulnerable commercial bank has less perceived value than money at a rock-solid commercial bank – and much less perceived value than cash or CBDC. In particular, ordinary households or small businesses that lack the capacity for financial planning and risk assessment might resort to the safe option on principle. In 2008 and 2010-2013, cash was relied upon (see subsection 2.3); in a world with CBDC, there is a second absolutely risk-free alternative.

**Bank run.** When depositors (retail and wholesale) withdraw their deposits at a high pace, this is referred to as a bank run. Therefore, one could say that all three scenarios describe a bank run, since runs are a permanent risk in this analysis – even though the speed of withdrawal was not discussed. In this context, the risks presented will not occur only when the previous scenario is fully reached. In reality, bank failures can occur much earlier than in the third scenario, and then a CB that is not subject to ceteris paribus clauses, must decide how much money to make available to a commercial bank on the brink of insolvency.

Table 2 provides an overview of the three scenarios.

| *Impact on …*<br>*Risks* | **commercial bank funding capabilities** | **commercial banks' business model** | **financial services industry** |
|---|---|---|---|
| **High-cost funding risk** | Funding reduced to longer-term liabilities and equity | Profitability decreases; shorter balance sheet; M&A to remain cost competitive | Banks lose loans & deposits; some LT liabilities may expand; market concentration increases |
| **Disintermediation risk** | Funding reduced to equity | Debt / equity ratio drops to zero; M&A with PSPs partially lead to the consolidation of liabilities | Distinct commercial bank disintermediation; more market concentration |
| **Solvency risk** | Not applicable | Resolution of commercial bank; digital bank run | What financial innovations would be used to fund businesses? |

**Table 2: Three scenarios for the future of commercial bank funding**

**Increased instability.** In addition to the three risks mentioned above, an increased stability risk for commercial banks appears again and again in the scenarios, be it in the attempt to use assets more profitably, to tap the wholesale markets more strongly, or in the speed of digital bank runs.

**The role of central banks.** Overall, the scenario analysis helps to better understand how an unconstrained substitution between commercial bank deposits and rCBDCs would lead to a fundamental redesign of the structure and scope of bank intermediation, and why CBs are reluctant to introduce rCBDCs. This is not a simple portfolio reallocation of money by households and non-financial businesses, this could be a run on the banking system. The main argument against issuing rCBDCs is that CBs should not compete with commercial banks. After all, the role of CBs is to supervise and provide liquidity to commercial banks. In other words, a CB follows the maxim of not disintermediating the banks.

Therefore, how can a CB introduce CBDC without derailing the commercial banks? The simple answer is to build trust in the commercial bank deposits. This is easier said than done. How can ordinary households be convinced that commercial bank deposits are at least as safe as CB money? So far, no one knows. Perhaps other features of commercial bank money can compensate for the bundle of security and trust? But if a CBDC is to be so unattractive that it does not have the potential to subvert the commercial banking system, the question of the purpose of a CBDC arises somewhere.

**Limitations and caps.** All of those scenario risks can be contained or nearly eliminated by restraints and ceilings. In the case of the Sand Dollar, excess funds must be transferred to the linked deposit accounts of domestic financial institutions. The governor of the CBOB clearly remarks, “We have not designed our CBDC as a substitute for deposit or equivalent assets in the banking system” [34].

These limitations and caps could theoretically be extended from the Bahamian version of household and corporate account restrictions and general ledger monitoring to restrictive conditions at the CB itself. Ultimately, that would amount to a restriction on convertibility and would massively reduce consumer trust. If a CBDC is to replace cash, as the CBDC definition in this paper suggests, an exclusive conversion of cash to the national CBDC may be worth considering. But how would you explain to a Bahamian that he or she can convert cash to CBDC, but not by transfer from a commercial bank account? Bahamians will be quick to notice they can simply withdraw cash (convertible to CBDC) from an ATM, maybe resulting in a bank run. Any risk of currency convertibility invites circumvention. Or, say, if the limit refers to a national total amount of CBDC, however defined: How would you explain to a Bahamian that he or she cannot deposit into his or her rCBDC eWallet because the money supply at the CB has reached its ceiling? Money depends on trust, see subsection 2.3.

**Fees.** A final thought on limitation would be fees, but fees for an official CB payment instrument that has legal tender status and is intended to replace cash seem outlandish. Government fees for a legal tender would significantly damage trust in this payment instrument.

# 5. The trade-off between financial privacy and tracing illicit payments

This section briefly discusses which regulations favor privacy, how it is undermined by laws and technical design, how the design of CBDCs addresses it, and finally summarizes the findings.

## 5.1 Protection of privacy

Financial privacy refers to the fact that the disclosure of financial data is prohibited in a country or internationally. Data protection and bank secrecy (in effect bank-client confidentiality) are enshrined in national and international law and are intended to protect clients from investigations, for instance by their own government. Privacy protection can include various elements, such as personal data (like identity), transaction data (like date and amount of payments, or the ledger itself), or other data (like account balances, online identifiers, keys etc.).

There are understandable reasons for wanting anonymous, untraceable transactions, which are perfectly legitimate, such as the finality of payments (e.g., a Bahamian merchant does not want to be accused a few days or weeks later by a foreign tourist that the goods purchased were defective) or the discomfort of payments (e.g., a customer buys perfectly legal goods, such as a bottle of bed bug spray). These examples are for illustrative purposes only and are by no means exhaustive. They are intended to show that the desire for financial privacy also exists outside of illegal activities.

Privacy protection is considered a key factor in the success of cryptocurrencies [46]. As some CBs devise CBDCs to combat competition from cryptocurrencies [1, 20, 47], the issue of consumer trust in privacy becomes crucial, since many users assume that cryptocurrencies can guarantee anonymity. While a token-based CBDC in a two-tier model (see [48] for an example) could provide anonymity to the CB, an account directly with the CB certainly does not.

Experts see things very differently. If anonymity refers to transaction data, then the open ledger of cryptocurrencies does not guarantee anonymity. And passive and active analysis of crypto-assets such as Bitcoin can completely de-anonymize individual users (personal data), but at great expense (see [49–51] for examples).

Figure 4 illustrates the level of privacy protection related to the general ledger for users in different payment systems, as assessed by [20].

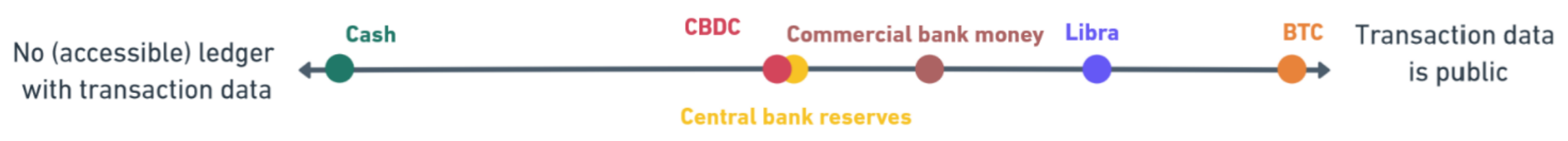

**Figure 4: The degree of privacy for users (source: [20], p. 34)**

Libra is a stablecoin (now called Diem), BTC is an abbreviation for Bitcoin. Cash has a greater advantage here than any other means of payment, as cash can actually be used without records.

Most cryptocurrencies treat privacy as an end in itself. Bitcoin's privacy weakness shown above has spawned services that allow transactions to be processed through a third party. These are called mixers because they aim to hide one transaction in a large number of unrelated transactions. However, anonymity includes not only the transactions, as shown in Figure 4, but also the identity of the payer and the payee. Since there are now many thousands of crypto-assets and each of them has different privacy and anonymity properties than others, this can lead to some confusion (see [46, 52] for surveys on anonymity and privacy in various crypto-assets). In any case, it is fair to say that cryptocurrencies have at least a major perceived advantage when it comes to privacy. Technically, researchers are getting closer and closer to cash-like privacy with ever new concepts, that include elements from onboarding to general ledger entry (the latter is irreplaceable to avoid double-spending).

## 5.2 Regulations and measures against privacy

Access to financial services without government control enables the hiding of proceeds from criminal undertakings (e.g., corruption), the financing of illegal activities (e.g., terrorism), and the evasion of taxes and regulations [53, 54].

While the level of data protection varies according to national legislation, the primary purpose of certain national regulations is to ensure that financial institutions keep records of transactions and report them to the authorities when required. Anti-money laundering (AML), combating the financing of terrorism (CFT), and Anti-Tax Avoidance (ATA) requirements aim to deter and detect illegal activities. International standards support or even drive this prioritization. The FATF has made the anonymity of virtual assets a "red flag indicator" for suspicious activity [54]. Indeed, the lifting of bank secrecy is enshrined in the most important international documents [55]. From a law enforcement perspective, data disclosure/transfer is seen as a legal tool, and data privacy is completely circumvented, nationally as well as internationally (see [17, 39, 49, 51] for examples; [53] for a well-known case of U.S. tax compliance; [56] for a comparison of US and EU legal frameworks on data protection in the field of law enforcement).

### 5.3 Privacy in CBDC design and the Bahamian Sand Dollar

AML/CFT and ATA requirements are not a core objective of a CBDC, but CBs are expected to ensure that CBDCs meet these requirements (along with other regulatory expectations or disclosure requirements) like any other financial institution [40]. Although some degree of anonymity can be achieved, whether through laws, bank-client confidentiality or token-based technology, it is implausible that CBDCs will be, or even could be, completely anonymous like cash [14, 40].

Nevertheless, some degree of anonymity in CBDC design, such as lower hurdles for identity verification or no linkage to bank accounts, would promote ease of use, enable a more ubiquitous access, and address privacy concerns [25]. In short, privacy protections can strengthen adoption of a CBDC.

Privacy protection for CBDCs can be done in a number of ways. Prepaid cards or eWallets could enable almost complete anonymity. The European Central Bank has developed and tested the concept of "anonymity vouchers," in which the AML authority periodically issues an additional status on the token to each CBDC user [48]. These statuses allow the anonymous transfer of a limited amount of CBDC funds within a specified time period, with the user's identity and transaction history not visible to the CB or to anyone other than the user's selected counterparties [48].

Can cash-like anonymity be achieved for a CBDC? Probably not. Even if the legal framework allows anonymity for small amounts during certain time periods, these conditions must be technically enforced. The concealment of larger transfers of funds through the parallel use of multiple pseudonyms for smaller transfers of funds could not be tolerated in a CBDC. This in turn requires technical identification of the payer or payee to prevent circumvention of the conditions. This reasoning also shows that complete anonymity and caps are not feasible at the same time for a CBDC. In the European Central Bank's concept of anonymity vouchers, anonymity may be achieved for small amounts in predefined time periods, but a KYC process takes place beforehand.

Privacy protection and bank-client confidentiality are of great importance in The Bahamas. The Bahamas has the reputation of being one of the most notorious tax havens in the world [57], a history of piracy, offshore scandals like the Bahamas Leaks [57] as well as an on-and-off relationship with various EU and FATF gray and blacklists due to AML/CFT/ATA deficiencies [58]. Unease or distrust about the security of a digital currency and its privacy is an issue in the Bahamas [27].

How far can this line of thinking take hold in the Sand Dollar? An important requirement for the Sand Dollar was that transactions should not be anonymous while at the same time protecting the confidentiality of the users [17, 27]. To facilitate access, revised AML guidelines in 2018 introduced streamlined customer due diligence standards that simplify identity and address verification requirements when establishing personal deposit accounts or accessing other AFI services [27, 59]. Requirements vary for low- and medium-value personal accounts [17, 27, 33]. Payment institutions may waive customer identification procedures for the small version of the eWallet. Nevertheless, the CBOB states in its annual report that the Sand Dollar is intended to help prevent money laundering and other illegal activities that are easier to commit with cash [30]. All transactions are linked to an AML/CFT engine, used by AFIs and owned by the CBOB, to ensure compliance (see [27] and Figure 2).

## 5.4 Results

CBDC design follows function, but design must also follow regulation. Customer identification and verification are just two elements of a broader KYC requirement that prevents true, comprehensive customer anonymity. They can be reduced or suspended altogether for smaller amounts, but the bottom line is that AML/CFT/ATA requirements and law enforcement will generally take precedence over data protection. This is illustrated in Figure 5.

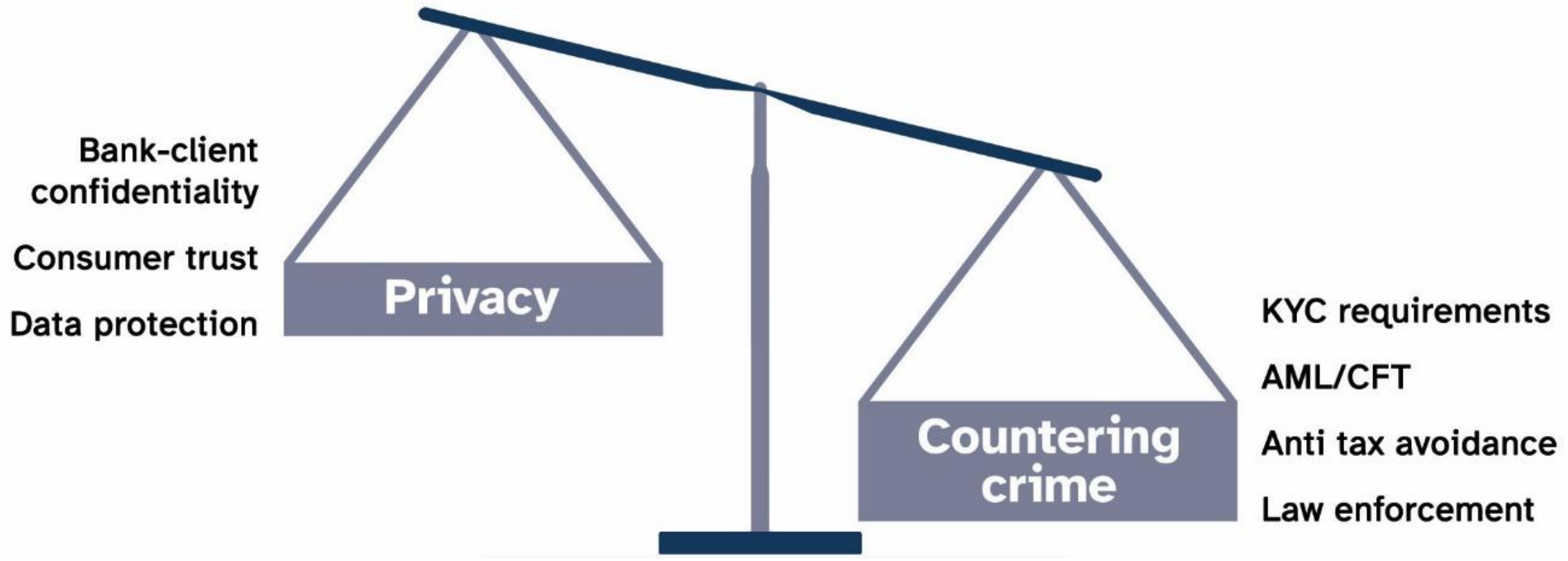

**Figure 5: Balancing financial privacy and measures against crime**

Thus, in competition with cryptocurrencies, a CB loses twice. Cryptocurrencies allow the almost anonymous transfer of unlimited amounts at any time. CBDCs, on the other hand, must be severely constrained by limits on holdings and transfers to avoid endangering the banking sector (see Section 4.4) and KYC/AML/CFT/ATA requirements prevent anonymity except for smaller amounts — while anonymity is considered an essential feature for the appeal of cryptocurrencies.

If protection of the banking sector and strict KYC/AML/CFT/ATA regulations prevent a rCBDC from replicating key cryptocurrency features such as near-cash anonymity and high-value transactions, what is left? In economies with a powerful banking sector, rCBDCs could be introduced as a possible substitute for cash in small-value, almost anonymous transactions. This brings us exactly to the design of the Sand Dollar.

# 6. Conclusion and perspectives

The fast-growing, market-driven demand for cryptocurrencies worries CBs, as monetary policy could be completely undermined. This is prompting many to contemplate CBDCs.

The Bahamian Sand Dollar is a striking example of a rCBDC. It is the first real-world example, it was launched in an offshore center known as a notorious tax haven, and many of its features incorporate solutions to currently theoretical problems. The Sand Dollar indicates that the use of restrictions and caps may be the new standard of a regulatory framework for rCBDCs if bank disintermediation is to be prevented.

Cryptocurrencies are (perceived to be) very anonymous. Conversely, an official currency such as a CBDC must comply with various KYC and record-keeping requirements, even in a tax haven like The Bahamas, and is therefore destined for less anonymity, although transactions involving small amounts could achieve significantly more anonymity than larger payments.

Some CBs want their CBDCs to be a game changer for cryptocurrencies, but not for the role and mission of CBs. This presents rCBDCs with the impossible task of keeping up with private cryptocurrencies and ideally pushing the latter back, but at the same time, limits and caps as well as less privacy will ensure that a rCBDC cannot gain much relevance. Therefore, it is likely that the next early movers in the field of CBDCs will either be motivated by overarching goals not considered in this paper, such as geopolitical ambitions and the avoidance of international sanctions, or will pursue other goals, such as banking the unbanked like in the case of The Bahamas — but not tackling crypto-assets.

I believe that researchers in the field of CBDCs, with their solid foundation in risk research and systems design, will contribute significantly to the study of rCBDCs, their underlying technologies, as well as the specific design. In particular, the topic of restraints and caps will be an exciting area of research. For CBDC attributes such as interest rates (positive or negative) and advanced features such as conditional payments based on DLT, the bulk of the work is still ahead of us.